\documentclass[aps,prc,twocolumn,showpacs,floatfix]{revtex4-1}

\usepackage{graphicx}% Include figure files
\usepackage{dcolumn}% Align table columns on decimal point
\usepackage{bm}% bold math
\usepackage{ulem}
\usepackage{color}

\newcommand{\ran}{($\alpha$,n)}
\newcommand{\rannull}{($\alpha$,n$_0$)}
\newcommand{\rani}{($\alpha$,n$_1$)}
\newcommand{\ranii}{($\alpha$,n$_2$)}
\newcommand{\raniii}{($\alpha$,n$_3$)}
\newcommand{\raniv}{($\alpha$,n$_4$)}
\newcommand{\ranj}{($\alpha$,n$_j$)}

\newcommand{\rap}{($\alpha$,$p$)}

\newcommand{\rgn}{($\gamma$,n)}

\newcommand{\rna}{(n,$\alpha$)}
\newcommand{\rnanull}{(n,$\alpha_0$)}

\newcommand{\ciii}{$^{13}$C}
\newcommand{\ovi}{$^{16}$O}

\newcommand{\spro}{$s$-process}

\begin{document}

\title{
  Revised Cross Section of the $^{13}$C($\alpha$,n)$^{16}$O reaction between 5
  and 8 MeV
}

\author{Peter Mohr}
\email[Email: ]{mohr@atomki.mta.hu}
\affiliation{
Diakonie-Klinikum, D-74523 Schw\"abisch Hall, Germany}
\affiliation{
Institute for Nuclear Research (Atomki), H-4001 Debrecen, Hungary}

\date{\today}

\begin{abstract}
As suggested in a Comment by Peters, Phys.\ Rev.\ C {\bf 96}, 029801 (2017), a
correction is applied to the $^{13}$C($\alpha$,n)$^{16}$O data of Harissopulos
{\it et al.}, Phys.\ Rev.\ C {\bf 72}, 062801(R) (2005). The correction refers
to the energy-dependent efficiency of the neutron detector and appears only
above the ($\alpha$,n$_1$) threshold of the $^{13}$C($\alpha$,n)$^{16}$O
reaction at about $E_\alpha \approx 5$ MeV. The corrected data are lower than
the original data by almost a factor of two. The correction method is verified
using recent neutron spectroscopy data and data from the reverse
$^{16}$O(n,$\alpha$)$^{13}$C reaction.
\end{abstract}

\maketitle

\section{Introduction}
\label{sec:intro}
The \ciii \ran \ovi\ reaction plays an important role in nuclear physics and
astrophysics. Many conventional nuclear physics experiments suffer from
background which is produced by the \ciii \ran \ovi\ reaction in carbon
buildup on the target although \ciii\ has only a small natural abundance of
about 1\%. In addition, the \ciii \ran \ovi\ reaction may be relevant as
radiogenic neutron background in underground laboratories (e.g.,
\cite{Coo18,Wes17,Mei09,Hea89,Fei68}). Here typical primary energies
$E_\alpha$ vary between about 5 and 9 MeV for the uranium and thorium decay
chains. As the \ran\ cross section decreases strongly towards low energies,
the relevant thick-target yield is essentially defined by the \ran\ cross
section close and slightly below the primary $E_\alpha$, i.e.\ between about 5
and 8 MeV. (All energies are given as laboratory energies
$E_{\alpha,{\rm{lab}}}$ or $E_{n,{\rm{lab}}}$ throughout this paper;
exceptions are explicitly stated.)

Unfortunately, this energy range above 5 MeV is not well-studied in
literature. Much work has been done to measure the \ciii \ran \ovi\ cross
section at very low energies. This energy range is important to determine the
stellar \ciii \ran \ovi\ reaction rate which defines the strength of the main
neutron source for the astrophysical \spro . The various experimental data
sets in the low MeV region \cite{Heil08,Har05,Bru93,Dro93,Bair73,Dav68,Sek67}
agree reasonably well, as e.g.\ discussed in the NACRE compilations
\cite{NACRE1,NACRE2} 
% 20180606
and in a recent review \cite{Cri18}.
% 20180606

The experimental data by Harissopulos {\it et al.}\ \cite{Har05} (hereafter:
Har05) extend the low MeV region up to about 8 MeV and are thus the only
experimental basis for the determination of radiogenic neutron yields from the
\ciii \ran \ovi\ reaction. However, these Har05 data have been questioned
severely in a recent Comment by Peters \cite{Pet17}. There it is stated that
``the actual cross section above 5 MeV could be almost 50\% lower than
reported by Harissopulos {\it et al.}'', and it is pointed out that there is a
problem with the neutron detection efficiency in the Har05 data. It is the aim
of the present study to further investigate the Har05 data above 5 MeV and to
provide a reliable correction to these experimental data.

\section{Re-Analysis of the Har05 data}
\label{sec:re}
The Har05 experiment used a $4 \pi$ thermal $^3$He neutron detector, embedded
in a cylindric polyethylene moderator. The determination of the neutron
efficiency $\eta$ for such a detector is a complicated problem because $\eta$
depends on the neutron energy. However, this information is lost because of
the thermalization of the neutrons in the moderator. It is worth noting that
similar problems with the neutron efficiency have been identified in a series
of \rgn\ experiments, performed at Livermore and Saclay; a correction to these
\rgn\ data was recently provided (e.g., \cite{Var17}). The present study
follows the idea of \cite{Var17} to provide improved data from a combination
of experimental and theoretical information.

In Har05, the neutron efficiency $\eta$ was determined as a function of the
neutron energy $E_n$ (in MeV) in their Eq.~(1) from 2 to 9 MeV. It is stated
that $\eta$ varies between 31\% at $E_\alpha = 0.8$ MeV and 16\% at $E_\alpha
= 8.0$ MeV. As pointed out by Peters \cite{Pet17}, the low efficiency $\eta$
at $E_\alpha = 8$ MeV indicates that Har05 assumed that the \ciii \ran
\ovi\ reaction is governed by the \rannull\ channel, leading to relatively
high neutron energies. However, slightly above $E_\alpha \approx 5$ MeV the
\rani , \ranii , \raniii , and \raniv\ channels open, and depending on the
branching, the average neutron energy $E_n$ is significantly lower and the
effective neutron detection efficiency $\eta_{\rm{eff}}$ is significantly
higher than assumed in Har05. Thus, instead of using the efficiency $\eta_0$
for the \rannull\ channel, an effective efficiency
\begin{equation}
\eta_{\rm{eff}} = \sum_{j=0}^{4} \, b_j(E_\alpha) \, \eta_j(E_{n,j})
\label{eq:eta_eff}
\end{equation}
has to be used where the $b_j$ are the neutron branchings of the
\ranj\ channel at a given $E_\alpha$, and the $\eta_j$ are the energy-dependent
detection efficiencies for neutrons from the \ranj\ channel. For the energy
range under study in Har05, the sum in Eq.~(\ref{eq:eta_eff}) runs over the
$^{16}$O $0^+$ ground state ($j=0$) and the excited states at $E_x = 6049$ keV
($0^+$), 6130 keV ($3^-$), 6917 keV ($2^+$), and 7117 keV ($1^-$).

Finally, this leads to a correction factor $f_{\rm{corr}}$ for the Har95 cross
section data:
\begin{equation}
f_{\rm{corr}} = \frac{\eta_0}{\eta_{\rm{eff}}}
\label{eq:f_corr}
\end{equation}
Obviously, the correction factor is $f_{\rm{corr}} = 1.0$ for energies below 5
MeV, and thus the agreement of the Har05 data with other literature data at
low energies is not affected by the present correction. For a vanishing
\rannull\ contribution (and thus low neutron energies around $E_\alpha \approx
8$ MeV) the correction factor will approach its lower limit $f_{\rm{corr}}
\approx 0.5$ which results from the given efficiency limits of 31\% at low and
16\% at high neutron energies in Eq.~(1) of Har05.

\begin{figure}[htb]
\includegraphics[bbllx=35,bblly=20,bburx=470,bbury=440,width=0.95\columnwidth,clip=]{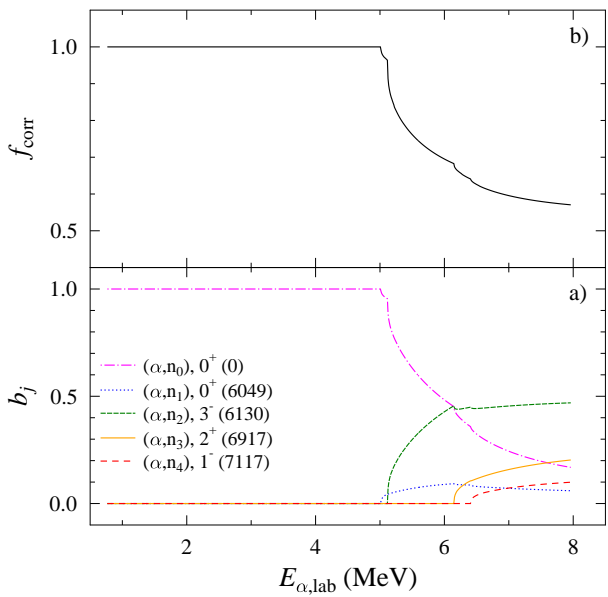}
\caption{
\label{fig:branch}
(Color online)
Branching ratios $b_j$ for the \ranj\ channels of the \ciii \ran
\ovi\ reaction as a function of energy $E_\alpha$ (lower part a) and resulting
correction factor $f_{\rm{corr}}$ from Eq.~(\ref{eq:f_corr}) for the cross
sections of Har05 (upper part b). 
}
\end{figure}

The present study uses the TALYS code \cite{TALYS,TALYS2} to calculate the
branching ratios $b_j$ of the \ranj\ channels. Of course, such a statistical
model approach can only be valid on average, and individual resonances in the
\ciii \ran \ovi\ reaction may show a completely different decay branching. But
it has been shown recently that a careful selection of TALYS parameters allows
to reproduce \ran\ cross sections for intermediate \cite{Mohr15,Tal18} and even
light nuclei \cite{Mohr17}, at least at energies $E_\alpha$ above a few
MeV. The calculated branching ratios $b_j$ as a function of energy $E_\alpha$
are shown in Fig.~\ref{fig:branch}. The correction factor $f_{\rm{corr}}$ is
then calculated from Eqs.~(\ref{eq:eta_eff}) and (\ref{eq:f_corr}) using the
energy-dependent efficiencies $\eta_j$ from Eq.~(1) of Har05 and the neutron
energies $E_{n,j}$ of the \ranj\ channels from reaction
kinematics. $f_{\rm{corr}}$ is also shown in Fig.~\ref{fig:branch}. All
numbers (Har05 cross sections, calculated branching ratios $b_j$,
efficiencies $\eta_0$ and $\eta_{\rm{eff}}$, correction factor $f_{\rm{corr}}$,
and the corrected cross sections) are provided as Supplemental Material to
this study \cite{Suppl}.

\begin{figure}[htb]
\includegraphics[bbllx=30,bblly=10,bburx=515,bbury=485,width=0.95\columnwidth,clip=]{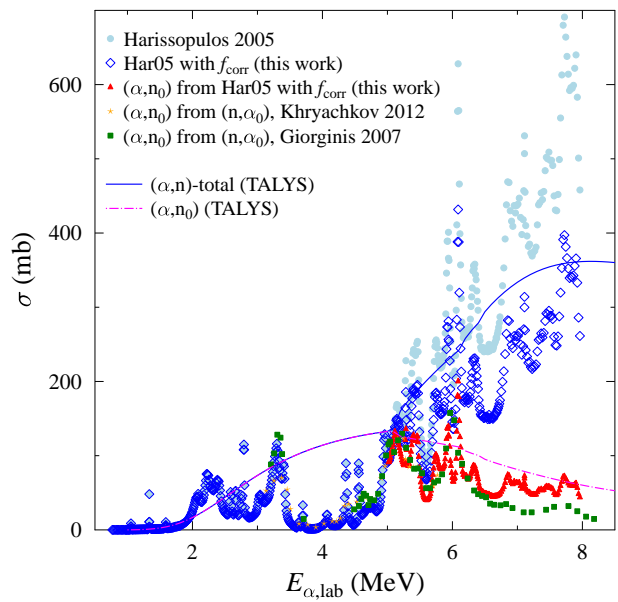}
\caption{
\label{fig:sigma}
(Color online)
Cross section of the \ciii \ran \ovi\ reaction. The corrected data (blue
diamonds) are significantly lower than the original Har05 data (lightblue
dots) for energies above the opening of the \rani\ channel at $E_\alpha
\approx 5$ MeV. The new estimated \rannull\ cross sections (red triangles) are
close to the results which are obtained from the reverse \ovi \rnanull
\ciii\ reaction (orange stars and green squares). Further discussion see text.
}
\end{figure}

The original cross sections of Har05 are shown in Fig.~\ref{fig:sigma} as
dots; the larger diamonds show the corrected data using $f_{\rm{corr}}$ from
Eq.~(\ref{eq:f_corr}) and Fig.~\ref{fig:branch}. Further details of
Fig.~\ref{fig:sigma} are discussed in the following Sect.~\ref{sec:disc}.

\section{Discussion}
\label{sec:disc}
Up to now, a statistical model calculation (using TALYS) was applied to
correct the experimental data of Har05. Fortunately, there are two ways to
verify the calculations and the applied correction factor $f_{\rm{corr}}$.

The first check uses the recently measured branching ratios $b_j$ by Febbraro
{\it et al.}\ \cite{Feb15}. Here a deuterated scintillator was used for
neutron spectroscopy, and it was possible to unfold the light response of the
scintillator to derive the neutron energies in the \ciii \ran \ovi\ reaction
at $E_\alpha = 7.5$ MeV (see Fig.~8 of \cite{Feb15}). It is found that the
\ranii\ channel dominates which populates the $3^-$ state in \ovi\ at 6130
keV. The \rannull\ ground state and \raniii\ $2^+$ (6917 keV) contributions
are about a factor of four smaller. Although no absolute efficiency
calibration was applied in \cite{Feb15}, the TALYS calculation nicely
reproduces the trend with a dominating \ranii\ channel (46\%), weaker
\rannull\ (20\%) and \raniii\ (19\%) channels, and minor contributions from
the \rani\ (6\%) and \raniv\ (9\%) channels at $E_\alpha = 7.5$ MeV.

The measured branching ratios $b_j$ of \cite{Feb15} clearly exclude the
assumption in Har05 that the \rannull\ channel is dominating, and it results
that the neutron energies are much lower than assumed in Har05. Consequently,
the correction factor $f_{\rm{corr}}$ in Eq.~(\ref{eq:f_corr}) and
Fig.~\ref{fig:branch} is confirmed.

A second test can be made using experimental data from the reverse \ovi \rna
\ciii\ reaction. The \ciii \rannull \ovi $_{\rm{g.s.}}$ cross section is
directly related to the \ovi \rnanull \ciii $_{\rm{g.s.}}$ cross section by
the reciprocity theorem. The relevant energy range is covered by the
\rnanull\ data by Khryachkov {\it et al.}\ \cite{Khr12} and Giorginis {\it et
  al.}\ \cite{Gio07} (as provided by EXFOR \cite{EXFOR}, including a
correction to the presented data in Fig.~8 of \cite{Gio07}). After conversion
from \rnanull\ cross sections to \rannull\ cross sections, the data of
\cite{Khr12} and \cite{Gio07} are also included in Fig.~\ref{fig:sigma}.

As expected, at low energies below the opening of the \rani\ channel, the
converted \rnanull\ data agree well with the \rannull\ data of Har05. However,
at higher energies the converted \rnanull\ data are significantly lower than
the Har05 data, reaching a discrepancy up to about one order of magnitude at
energies around $7-8$ MeV. This finding again invalidates the approach by
Har05 that the \rannull\ channel is dominating.

An attempt is made to estimate the \rannull\ cross section from the corrected
Har05 data and the calculated ground state branching $b_0$ (red triangles in
Fig.~\ref{fig:sigma}). These estimated \rannull\ data are close to the
converted \rnanull\ data of \cite{Gio07} (green squares). At energies above 6
MeV, the estimated data are still slightly higher than the converted
\rnanull\ data; this can be interpreted as evidence that most of the
resonances in the \ran\ data at higher energies preferentially decay to
excited states in \ovi , but not to the \ovi\ ground state.

Both above methods of verification confirm that the TALYS calculation of the
ground state branching is realistic with a trend that the real ground state
branching may be even lower than the calculated $30\% - 15\%$ above 6.5
MeV. Thus, it becomes obvious that a correction to the Har05 data has to be
applied where a ground state branching $b_0 = 1.0$ was assumed. A correction
factor $f_{\rm{corr}} \approx 0.65 - 0.55$ is determined above $E_\alpha
\approx 6.5$ MeV, with a lower limit of about 0.5 (for a vanishing ground
state branching $b_0$ and thus low neutron energies $E_n$). This leads to an
uncertainty of the correction factor $f_{\rm{corr}}$ of the order of $10\% -
20\%$. This result is almost independent of details of the $b_j$ ($j \ne 0$)
branching ratios towards excited states in \ovi\ because only the
\rannull\ channel leads to neutrons with relatively high energies.
The uncertainty of $\eta_{\rm{eff}}$ and $f_{\rm{corr}}$  may be somewhat
larger close above the respective \ranj\ thresholds where the neutron emission
in the laboratory is kinematically focused to forward directions.

The uncertainty of the correction factor $f_{\rm{corr}}$ is explained in more
detail for the energies of 6 MeV and 7.5 MeV, i.e.\ relatively close above the
threshold of the \rani\ channel and at the energy of the new experimental data
of \cite{Feb15}. At 6 MeV, the calculated branching ratios are $b_0 = 0.48$,
$b_1 = 0.09$, and $b_2 = 0.43$, leading to an effective efficiency
$\eta_{\rm{eff}} = 29.4$\% instead of $\eta_0 = 20.4$\%. The uncertainty of
the calculated ground state branching $b_0$ is carefully assumed with a factor
of two. This leads to an upper limit of $b_0 \approx 1$ and to a lower limit
$b_0 = 0.24$. Obviously, for the upper limit of $b_0$ I find $\eta_{\rm{eff}}
= 21.0\% \approx \eta_0$. The lower limit of $b_0$ results in an increased
$\eta_{\rm{eff}} = 33.6$\%. Consequently, $f_{\rm{corr}} =
0.695^{+0.278}_{-0.087}$. At 7.5 MeV, the corresponding numbers are $b_0 =
0.20$, $b_1 = 0.06$, $b_2 = 0.46$, $b_3 = 0.19$, and $b_4 = 0.09$, leading
$\eta_{\rm{eff}} = 31.4$\% instead of $\eta_0 = 18.2$\%. The upper and lower
limits of $b_0$ (again assuming a factor of two uncertainty for $b_0$) result
in a range of $\eta_{\rm{eff}}$ between 28.4\% and 33.0\% and $f_{\rm{corr}} =
0.580^{+0.062}_{-0.027}$. Summarizing, even the assumed significant
uncertainty of a factor of 2 for the ground state branching $b_0$ translates
to a typical uncertainty of the correction factor $f_{\rm{corr}}$ of the order
of $10-20$\%. Note that this result is almost independent on the detailed
branching towards the 4 excited states because the excitation energies are
within about 1 MeV, and thus the neutron energies are low and very similar for
all branchings $b_1$, $b_2$, $b_3$, and $b_4$.

Of course, these uncertainties should be considered as average uncertainties,
i.e.\ uncertainties of the average cross sections over a significant
energy interval. Individual resonances (as visible in Fig.~\ref{fig:sigma})
may show a completely different branching than calculated by TALYS. In
the extreme case of a resonance with a full ground state branching $b_0 =
1.0$, the correction factor $f_{\rm{corr}} = 1.0$ remains unity within the
energy interval of this resonance. Thus it is not meaningful to provide
uncertainties for each data point of the corrected Har05 data. Instead, an
overall uncertainty of about 15\% is recommended for yield calculations which
average over a sufficiently wide energy interval of at least a few hundred keV.

In principle, the experimental approach of Har05 can also be used to provide
at least a rough estimate of the neutron energy via the so-called ``ring
ratio'': the ratio of the neutron yields in the outer and inner ring of the
Har05 neutron detector depends on the neutron energy. Unfortunately, the
experimental setup of Har05 used only one ADC for the sum signal of all
neutron detectors, and thus no ring ratio can be provided from the Har05
experiment \cite{Har18}.

It is also interesting to see that in general the statistical model
calculation provides a reasonable agreement (on average) with the experimental
\ciii \ran \ovi\ data (see Fig.~\ref{fig:sigma}). However, the calculation
clearly overestimates the experimental data around $E_\alpha \approx 3.5 - 5$
MeV. This energy interval shows a relatively small number of resonances,
compared to lower and higher energies. It is not surprising that the agreement
between the statistical model calculation and the experimental data becomes
better in regions with a higher number of resonances, but even at the highest
energies under study between 6 and 8 MeV the calculation is slightly higher
than the average of the experimental data. The overestimation of the
experimental cross sections in the statistical model does not affect the
correction factor $f_{\rm{corr}}$ which depends only on the calculated
branching ratios $b_j$. Interestingly, a similar overestimation for the TALYS
calculation is also found for new preliminary data of the $^{13}$N\rap
\ovi\ mirror reaction \cite{Tal18b}.

Finally, a brief comparison to R-matrix fits from literature
\cite{Heil08,Kun14} is provided. The fit by Heil {\it et al.}\ \cite{Heil08}
did not include the Har05 data, but was constrained by \rna\ data up to
neutron energies of 8.5 MeV. Above about $E_\alpha = 5$ MeV, the fit of the
\ran\ data in Fig.~17 of \cite{Heil08} is lower than the Har05 data whereas
the fit agrees with the Har05 data at lower energies. This result is
consistent with the findings of the present study. The later study by Kunieda
{\it et al.}\ \cite{Kun14} uses the Har05 data for fitting. But unfortunately
this study focuses on the low-energy region with $E_\alpha < 4.6$ MeV, and no
conclusion can be drawn from \cite{Kun14} for the energy range under study in
this work.

\section{Conclusions}
\label{sec:conc}
The \ciii \ran \ovi\ data of Harissopulos {\it et al.}\ \cite{Har05} cover a
wide energy range from about 0.8 MeV to 8 MeV. At low energies below the
opening of the \rani\ channel at about 5 MeV, these data agree well with
various literature data. The cross sections between 5 MeV and 8 MeV are
important for the estimate of radiogenic neutron background in low-background
environments like underground laboratories. In this energy range experimental
data are rare, and the experimental data by Harissopulos {\it et al.}\ have
been questioned in a Comment by Peters \cite{Pet17}.

Following the criticism by Peters, the present study provides a correction to
the experimental data which is based on an improved determination of the
neutron detection efficiency $\eta_{\rm{eff}}$. Whereas the original study of
Harissopulos {\it et al.}\ assumed a dominating \rannull\ ground state
contribution (with resulting high neutron energies and low detection
efficiency), the present work finds a dominating \ranii\ channel, populating
the $3^-$ state in \ovi\ (with resulting lower neutron energies and higher
detection efficiency). The derived correction factor $f_{\rm{corr}}$ decreases
from unity at the opening of the \rani\ channel at $E_\alpha \approx 5$ MeV
down to about 0.55 at $E_\alpha \approx 8$ MeV. The applied method and the
resulting $f_{\rm{corr}}$ are validated by further studies which are based on
recent neutron spectroscopy data \cite{Feb15} and on data from the reverse
\ovi \rna \ciii\ reaction \cite{Khr12,Gio07}. The corrected \ciii \ran
\ovi\ cross sections are reliable with uncertainties of about 15\%. A further
reduction of uncertainties requires new experiments which should use improved
neutron detectors, either with spectroscopic properties \cite{Feb15} or with
an almost flat detection efficiency (as e.g.\ suggested in \cite{Uts17}).

\acknowledgments
I thank R. Talwar and K.\ E.\ Rehm for motivating this study, and
S.\ Harissopulos and H.-W.\ Becker for encouraging discussions.
This work was supported by NKFIH (K108459 and K120666).

\end{document}